\documentclass[10pt,journal,twocolumn,twoside]{IEEEtran} 
\usepackage{graphicx}
\usepackage{epstopdf}
\usepackage{float}
\usepackage{algorithmic}
\usepackage{array}
\usepackage{amsmath}
\usepackage{amssymb}
\usepackage{mdwmath}
\usepackage{mdwtab}
\usepackage{booktabs}
\usepackage{eqparbox}
\usepackage{stfloats}
\usepackage{fixltx2e}
\usepackage{hyperref}
\usepackage{cleveref}
\hypersetup{hypertex=true,
            colorlinks=true,
            linkcolor=blue,
            anchorcolor=blue,
            citecolor=blue}
\usepackage{cases} 
\usepackage{subfigure}
\usepackage{flushend}
\usepackage{upgreek}

\usepackage{xcolor}
\usepackage{makecell}
\usepackage{amsmath}
\usepackage[boxed,ruled,commentsnumbered]{algorithm2e}
\usepackage{url}
\usepackage{cite}
\ifCLASSOPTIONcompsoc
\usepackage[caption=false,font=normalsize,labelfont=sf,textfont=sf]{subfig}
\else
\fi
\allowdisplaybreaks[4]

\makeatletter

\renewcommand*{\@opargbegintheorem}[3]{\trivlist
      \item[\hskip \labelsep{\bfseries #1\ #2}] \textbf{(#3):}\ }
\makeatother

\begin{document}
\title
{Learning Multi-Rate Task-Oriented Communications Over Symmetric Discrete Memoryless Channels}
\author{Anbang Zhang, and Shuaishuai Guo,~\IEEEmembership{Senior Member, IEEE}
\thanks{The work is supported in part by the National Natural Science Foundation of
China under Grant 62171262; in part by Shandong Provincial Natural Science
Foundation under Grant ZR2021YQ47; in part by the Taishan Young Scholar
under Grant tsqn201909043; in part by Major Scientific and Technological
Innovation Project of Shandong Province under Grant 2020CXGC010109. (\emph{*Corresponding author: Shuaishuai Guo})

Anbang Zhang, and Shuaishuai Guo are with School of Control Science and Engineering, Shandong University, Jinan 250061, China (e-mail: 202234946@mail.sdu.edu.cn, shuaishuai\_guo@sdu.edu.cn).

Code is available in \url{https://github.com/zab0613/MR-ToC}}

}

\maketitle


\begin{abstract} 

This letter introduces a multi-rate task-oriented communication (MR-ToC) framework. This framework dynamically adapts to variations in affordable data rate within the communication pipeline. It conceptualizes communication pipelines as symmetric, discrete, memoryless channels. We employ a progressive learning strategy to train the system, comprising a nested codebook for encoding and task inference. This configuration allows for the adjustment of multiple rate levels in response to evolving channel conditions. The results from our experiments show that this system not only supports edge inference across various coding levels but also excels in adapting to variable communication environments.

\end{abstract}


\section{Introduction} 


\IEEEPARstart{T}{raditional}  communication systems require accurate bit recovery at the receiver \cite{8723589}, which depends on good channel conditions and enough communication resources. They are designed as general-purpose pipelines for data transmission without special consideration for the nature of the current task or requirements. While widely used, this static model often fails to address needs in situations \cite{Shao2021LearningTC} where time sensitivity and data relevance are critical. Therefore,  prioritizing the transmission of task-relevant information attracts much research attention \cite{9955525}. The emergence of task-oriented communication (ToC) systems
is expected to be a major leap forward in disrupting the traditional paradigm in digital communication. This shift is particularly relevant to remote inference tasks, especially in the various industry domains belonging to the era of big data, where timely and accurate data analysis is critical to the decision-making process. 

Several paradigms have been proposed to reduce the amount of required information to perform the task, such as information bottleneck \cite{Shao2021LearningTC}, feature compression \cite{9261169}, semantics-aware coding \cite{9475174}. To this end, the authors in \cite{9837474}, utilize the distributed information bottleneck (DIB) framework to develop task-oriented communication strategies for multi-device cooperative inference. Based on the $\beta$-variational autoencoder ($\beta$-VAE), a practical explainable semantic communication system design \cite{10110357} is proposed to improve the transmission efficiency. In \cite{10483054}, the authors have proposed a utility-informativeness-security trade-off framework to deal with achieving better task performance while reducing a certain amount of information.

Although task-oriented end-to-end communication architectures based on learning-driven schemes have shown to be effective in saving communication bandwidth \cite{10159007}, the practical application of these systems faces significant obstacles due to the dynamic nature of current bit-pipeline infrastructure, which has dynamic affordable transmission rates. The challenges posed by varying data rates are not just technically troublesome but are fundamental barriers that can destroy the integrity and reliability of remote inference tasks. These challenges are further exacerbated by the inability of existing systems to adapt to changing communication pipeline conditions, resulting in an urgent need for more functional and resilient end-to-end communication frameworks. 

Motivated by this, we introduce an adaptive multi-rate ToC framework, named MR-ToC. We leverage an adaptive tuning scheme with the nested codebook for  MR-ToC, which can learn unified encoder and inference models through a progressive learning approach. 
The knowledge base utilized in our proposed framework is the nested codebook, which differs from the common knowledge base typically found in traditional semantic communication systems \cite{10454584}. The MR-ToC framework allows the use of a single encoder, a task inference module, and the designed nested codebook across various transmission rates. For instance, in a wireless channel with adaptive modulation, when the modulation order changes, the proposed MR-ToC scheme can be seamlessly applied without significant modifications. In contrast, existing ToC frameworks developed for fixed modulation orders would require adjustments to the encoder, task inference module, and codebook whenever the modulation order changes. This flexibility underscores the practical advantages of our approach in dynamic communication environments.

\section{System Model} 

In this section, we introduce the MR-ToC system in which transceiver utilizes a shared, learned nested codebook. Following this, we provide a detailed description of the problem pertaining to the multi-rate learning framework.

\subsection{Discrete Task-Oriented Communication System}

\begin{figure*}[htbp]
\centering
\includegraphics[width=1.9\columnwidth,height=0.75\columnwidth]{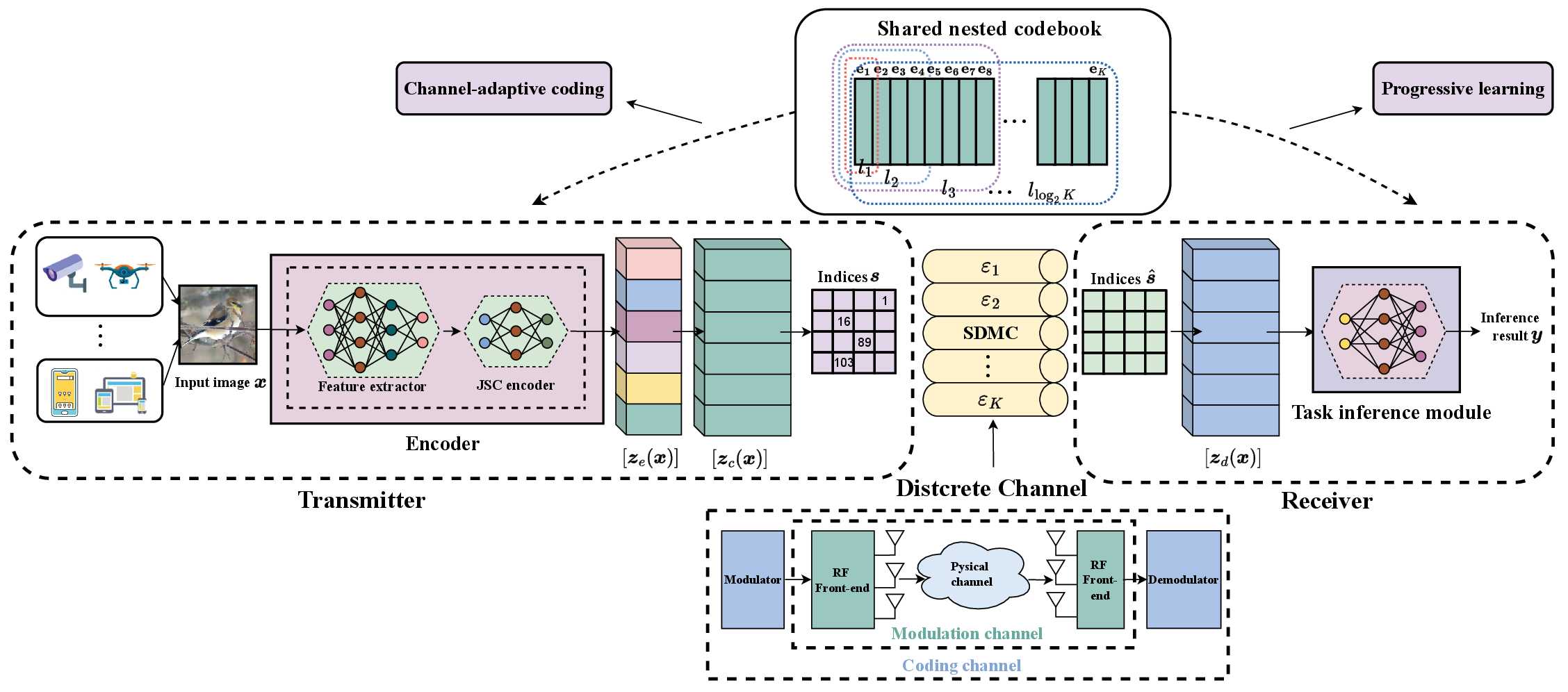}
\caption{Discrete multi-rate task-oriented semantic communication over multilevel symmetry discrete memoryless channels}
\label{fig1}
\end{figure*}

As illustrated in Fig. \ref{fig1} of the MR-ToC framework, there is a data source that generates the data $\boldsymbol{x}\in{\mathbb{R}^{N}}$ and its target variable $\mathbf{y}$ (i.e. label) at the transmitter side. A feature extractor first identifies the task-relevant feature from the raw input ${x}$, which is mapped to a continuous representation $\boldsymbol{z}_{e}(\boldsymbol{x})\in {\mathbb{R}^{N_f}}$ through joint source and channel (JSC) encoder, as
\begin{equation}\label{eq1}
\boldsymbol{z}_{e}(\boldsymbol{x}) = E(\boldsymbol{x} ; \pmb{\theta}),
\end{equation}
where ${E(\boldsymbol{x}; \pmb{\theta})}$ is a combination of the feature extractor and JSC encoder at the transmitter side with parameters $\pmb{\theta}$.

\emph{1) Discrete Codebook Mapping:} 
To be compatible with existing digital communication systems, we exploit vector quantization to discretize continuous encoder output. Let $Q=\left\{\mathbf{e}_{j}\in \mathbb{R}^{D}| j=1,2,\cdots,K\right\} $ represent the whole codebook, where $\mathbf{e}_{i}$ represents the $i$th codeword and ${K}$ is the number of codewords ($|Q| = {K}$).  Subsequently, the continous vector $\boldsymbol{z}_{e}(\boldsymbol{x})$ is decomposed into $M$ sub-vectors, each of which is then mapped to its closest codeword in $Q$. Moreover, the number of bits conveyed is $B=M \cdot \log _{2} K$. Thus, the discrete vector $\boldsymbol{z}_{c}(\boldsymbol{x})$ is shown by the nearest-neighbor search as
\setcounter{equation}{1}
\begin{equation}\label{eq2}
\boldsymbol{z}_{c,m}(\boldsymbol{x})=\arg \min _{\mathbf{e}_{j}}\left\|\boldsymbol{z}_{e,m}(\boldsymbol{x})-\mathbf{e}_{j}\right\|_{2}, \forall \mathbf{e}_{j}\in \mathcal{Z}_c.
\end{equation}
where $\boldsymbol{z}_{c,m}(\boldsymbol{x})$ and $\boldsymbol{z}_{e,m}(\boldsymbol{x})$ represent the $m$-th sub-vectors of $\boldsymbol{z}_{c}(\boldsymbol{x})$ and $\boldsymbol{z}_{e}(\boldsymbol{x})$, respectively. The feature/codewords indices are directly encoded and then transmitted.

Note that the codebook is obtained through joint training within an end-to-end learning framework. The continuous-to-discrete mappings pose a problem for learning due to its inherent non-differentiable nature. To deal with this issue, we adopt a straight-through estimator method in jointly training the encoder-inference modules while learning the codebook $Q$. Specifically, in the forward propagation, the nearest vector (i.e., $\boldsymbol{z}_{c}(\boldsymbol{x})$) is computed by (\ref{eq2}) and propagated. In backward propagation, since the “$\mathrm{arg~min}$” operation is non-differentiable, the gradient of the reconstructed error can not be computed and propagated. Instead, we approximate the forward propagation and the gradient by a straight-through estimation $\mathcal{L}_{\text{VQ}}$ \cite{NIPS2017_7a98af17} as
\begin{equation}\label{eq3}
\begin{aligned}
\mathcal{L}_{\text{VQ}}=\left\|\mathbf{sg}\left[\boldsymbol{z}_{e}(\boldsymbol{x})\right]-\mathbf{e}_{j}\right\|_{2}^{2}+\gamma\left\|\boldsymbol{z}_{e}(\boldsymbol{x})-\mathbf{sg}\left[\mathbf{e}_{j}\right]\right\|_{2}^{2},
\end{aligned}
\end{equation} 
where $\gamma$ is the hyper-parameter and $\mathbf{sg}[\cdot]$ represents the stop gradient operation. The first term is the VQ-loss, which moves the codebook vectors closer to the encoder outputs. The second term is the commitment loss, which causes the encoder outputs to be similar to the codebook vectors. And the hyper-parameter $\gamma > 0$ balances the influence of the commitment loss on $\mathcal{L}_{\text {VQ}}$. The stop gradient operation acts as an identity function at forward propagation and has zero partial derivatives, constraining its operand to be a non-updated constant.  

\emph{2) Symmetry Discrete Memoryless Channels:} To transmit the encoded discrete representation, we utilize the SDMC to model this transmission of discrete indices.
The effects of modulation/demodulation techniques, power control, and physical channel conditions are reflected in the
parameters of the discrete channels.
The SDMC model assumes that both channel inputs and outputs utilize the same symbol set. This model is represented by a probability transfer matrix:
\begin{equation}\label{eq4}
\mathcal{P}=\left[\begin{array}{cccc}
1-\varepsilon & \frac{\varepsilon}{r-1} & \cdots & \frac{\varepsilon}{r-1} \\
\frac{\varepsilon}{r-1} & 1-\varepsilon & \cdots & \frac{\varepsilon}{r-1} \\
\cdots & \cdots & \cdots & \cdots \\
\frac{\varepsilon}{r-1} & \frac{\varepsilon}{r-1} & \cdots & 1-\varepsilon
\end{array}\right]_{r \times r},
\end{equation}
where $\varepsilon$ denotes the error transmission probability to simulate the different channel conditions and $r$ represents the total number of used 
codewords, adapting to the codebook size. Thus, the potential transmission errors could lead to retrieving incorrect vectors, thereby impacting the inference task.

The corresponding codewords indices of discrete vector $\boldsymbol{z}_{c}(\boldsymbol{x})$ is mapped to the transmitted symbol. Let $\mathbf{s}$ represent the transmitted signal, then which is fed into the SDMC, and the detection at the task inference device can be expressed as $\hat{\boldsymbol{s}}$. The task inference device performs demapping by selecting the corresponding vectors $\boldsymbol{z}_{d}(\boldsymbol{x})$ in the discrete codebook according to the received symbols $\hat{\boldsymbol{s}}$. 

\emph{3) Task Interference:} After recovering the signal, $\boldsymbol{z}_{d}(\boldsymbol{x})$ is utilized by the task inference module for the remote inference task, which is shown as:
\begin{equation}\label{eq5}
\hat{y}=R(\boldsymbol{z}_{d}(\boldsymbol{x}) ; \pmb{\delta}),
\end{equation}
where $R(\boldsymbol{z}_{d}(\boldsymbol{x}) ; \pmb{\delta})$ is the task inference module with parameters $\pmb{\delta}$. 

In summary, the whole MR-ToC system can be modeled by a  probabilistic graphical model:
\begin{equation}\label{eq6}
Y \rightarrow X \rightarrow Z_{e} \rightarrow Z_{c} \rightarrow S \rightarrow \hat{S} \rightarrow Z_{d} \rightarrow \hat{X} \rightarrow \hat{Y}, 
\end{equation}
satisfying 
\begin{align}\label{eq6a}
&p(\hat{\boldsymbol{y}} | \boldsymbol{x})  = \\ &p_{\pmb{\delta}}(\hat{\boldsymbol{y}} | \boldsymbol{z}_{d}) 
p_d\left(\boldsymbol{z}_{d} | \hat{\boldsymbol{s}}\right) 
p_{\text{channel }}\left(\hat{\boldsymbol{s}} | \boldsymbol{s}\right) \nonumber 
p_m\left(\boldsymbol{s} | \boldsymbol{z}_{c} \right) 
p_e\left(\boldsymbol{z}_{c} | \boldsymbol{z}_{e}\right) 
p_{\pmb{\theta}}\left(\boldsymbol{z}_{e} | \boldsymbol{x}\right),
\end{align}
where $S$, $\hat{S}$, $Z_{e}$, $Z_{c}$, $Z_{d}$, $X$, $\hat{X}$ and $Y$ are random variables; $\boldsymbol{s}$, $\hat{\boldsymbol{s}}$, $\boldsymbol{z}_{e}$, $\boldsymbol{z}_{c}$, $\boldsymbol{z}_{d}$, $\boldsymbol{x}$, $\hat{\boldsymbol{x}}$, and $\mathbf{y}$ are their instances, respectively; and $p_{(\cdot)}(\cdot|\cdot)$ represents the transmission function.

\subsection{Problem Description}
Given that communication pipelines are inherently dynamic, resulting in fluctuating data rates, there is a pressing need to reevaluate traditional task-oriented communication systems. These conventional systems, which include encoders, codebooks, and task inference modules, are typically optimized for a fixed transmission rate. To effectively manage a variable-rate pipeline, it is necessary for the transmitter to incorporate an array of encoders, codebooks, and task inference modules, thereby allowing flexibility in response to changing conditions. This raises a critical question: Is it feasible to develop a universally adaptable system that can efficiently operate across all potential transmission rates?

Given an affordable data rate of $V_{bit}$ bit per second and a BER of $P_e$, to accomplish the transmission within a delay of $\tau_t$, we have to use the codebook size of $K_t$, expressed as
\begin{equation}\label{eq8}
K_t=\frac{V_{bit}\tau_t}{\log_2 M}.
\end{equation}
The error transmission probability of SDMC $\epsilon$ can be set as
\begin{equation}\label{eq9}
\epsilon= 1-(1-P_e)^{\log_2 K_t}.
\end{equation}
To develop an end-to-end learning architecture that caters to multi-rate requirements, we employ a nested codebook combined with a progressive learning strategy. In this letter, we utilize a classification task as a case study to illustrate our training methodology. The effectiveness of this approach is quantified by the accuracy of the predictions, denoted as $\hat{y}_{t}$.

\begin{algorithm}[t]
  \label{alg1}
  \KwIn{Dataset $\mathcal{D}$; SDMC {$\varepsilon$}; Encoder $\pmb{\theta}$; Inference module $\pmb{\delta}$; Parameters {$\eta_{l}$}, {$\gamma_{j}$}, and $K_{\max}$.}
  Initialize $Q_{0}=\emptyset$;

  \For{quantization level $l \in \{1, 2, \dots, \log_2 K_{\max}\}$}{
  Init $Q_l$ by adding $2^{l-1}$ random vectors to $Q_{l-1}$;
    
  \For{epoch $= 0, 1, \dots$}{
  Randomly divide $D$ into $P$ batches $\{D_p\}^P_{p=1}$;
        
  \For{$p = 1, \dots, P$}{
  \emph{Encoder:} $E( \mathcal{D} ; \pmb{\theta}) \rightarrow \boldsymbol{z}_{e}(\boldsymbol{x})$;
  
  \emph{Discretization:} $\boldsymbol{z}_{e}(\boldsymbol{x}) \rightarrow \boldsymbol{z}_{c}(\boldsymbol{x})$;
  
  \emph{SDMC:} Transmit codewords' indices $s$ over the channel;

  \emph{Inference:} $R( \boldsymbol{z}_{d}(\boldsymbol{x}) ; \pmb{\delta})\rightarrow \hat{y}$;
  
  Calculate the loss using $\mathcal{L}_{\text {MR-ToC}}^{(l)}$ in (\ref{eq12});
  
  Update $T( \cdot ; \pmb{\theta})$, $Q_l$, and $R( \cdot ; \pmb{\delta})$.}
    }
  }
  \KwOut{Trained $T( \cdot ; \pmb{\theta})$, $R( \cdot ; \pmb{\delta})$, and $Q_{\log_2 K_{\max}}$}
  \caption{MR-ToC Training Over SDMC}
\end{algorithm}

\section{Adaptive Coding in Progressive Learning} 
In this section, we introduce multi-rate transmission mechanism and discuss the progressive learning training strategy. 

\subsection{Multi-Rate Transmission Mechanism}
\emph{1)Prior Criteria:} To achieve a multi-rate transmission mechanism, we consider training MR-ToC with different codebook size $K_t$. There are two main points about the mechanism as: ($i$) prepare a codebook that can be applied at different data rate; ($ii$) ensure that the task inference module can reliably infer using received representations at different resolutions.

In detail, we utilize a large-size codebook that can be successively decomposed into sub-codebooks of small size \cite{10193526}, supporting multi-rate coding with a single codebook. The developed codebooks, including themselves, are a subset of valid codebooks. Thus, the codewords used for encoding the
data to $B_{i}$ bits are a subset of the codewords used for encoding to $B_{j}$ bits for every $B_{i}<B_{j}$. Based on large-size codebooks and their small-size subsets,  MR-ToC framework is trained progressively to improve task inference performance.

\emph{2) Channel-Aware Rate-Adaptive Codebook Design:} Let  ${K}_{\max}$ represent the largest codebook size corresponding to the codebook to support the maximal coding rate. For each $l=1, \ldots, \log _{2} K_{\max}$, we denote the corresponding codebook $Q_{l}$. Instead of setting the codebooks for each size individually, we set codebooks for various coding rates in a nested form $Q_{1} \subset Q_{2} \subset \cdots \subset Q_{\log _{2} K_{\max}}$.
At time $t$, the MR-ToC selects the coding level $l_{t}$  for which
the latency constraint holds as 
\begin{equation}\label{eq11}
l_{t}=\max \left\{l \in\left\{1, \ldots, \log _{2} K_{\max}\right\} \left\lvert\, \frac{M\cdot{l}}{V_{bit}} \leq \tau_{t}\right.\right\}.
\end{equation}

Then, the proposed MR-ToC framwork perform inference task through the codebook $Q_{l_{t}}$.

\begin{figure*}[!t]
        \centering
        \subfigure[Codebook $Q_{2}$.]{{\label{fig2a}}\includegraphics[width=0.245\linewidth]{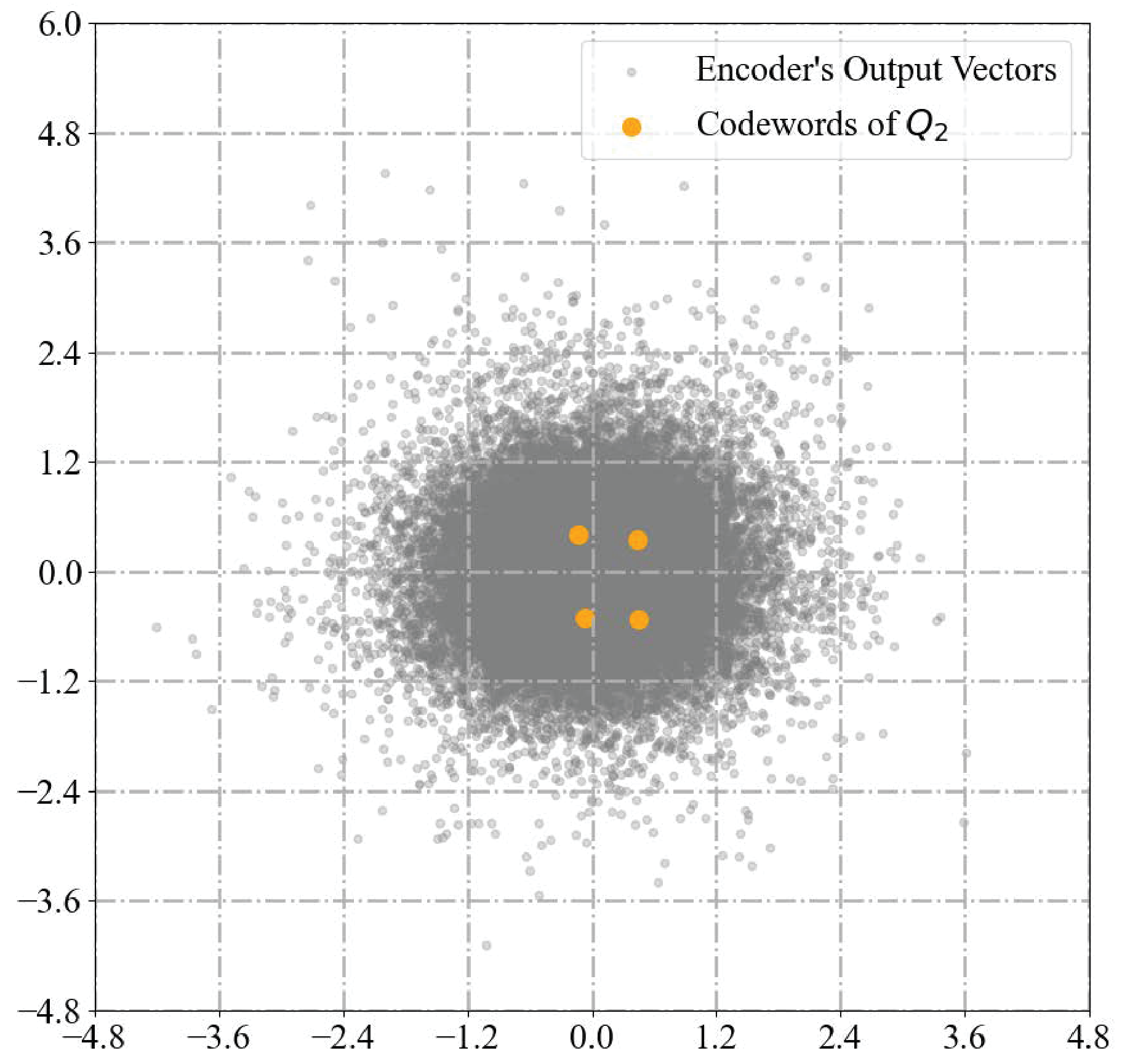}}
        \subfigure[Codebook $Q_{4}$.]{{\label{fig2b}}\includegraphics[width=0.245\linewidth]{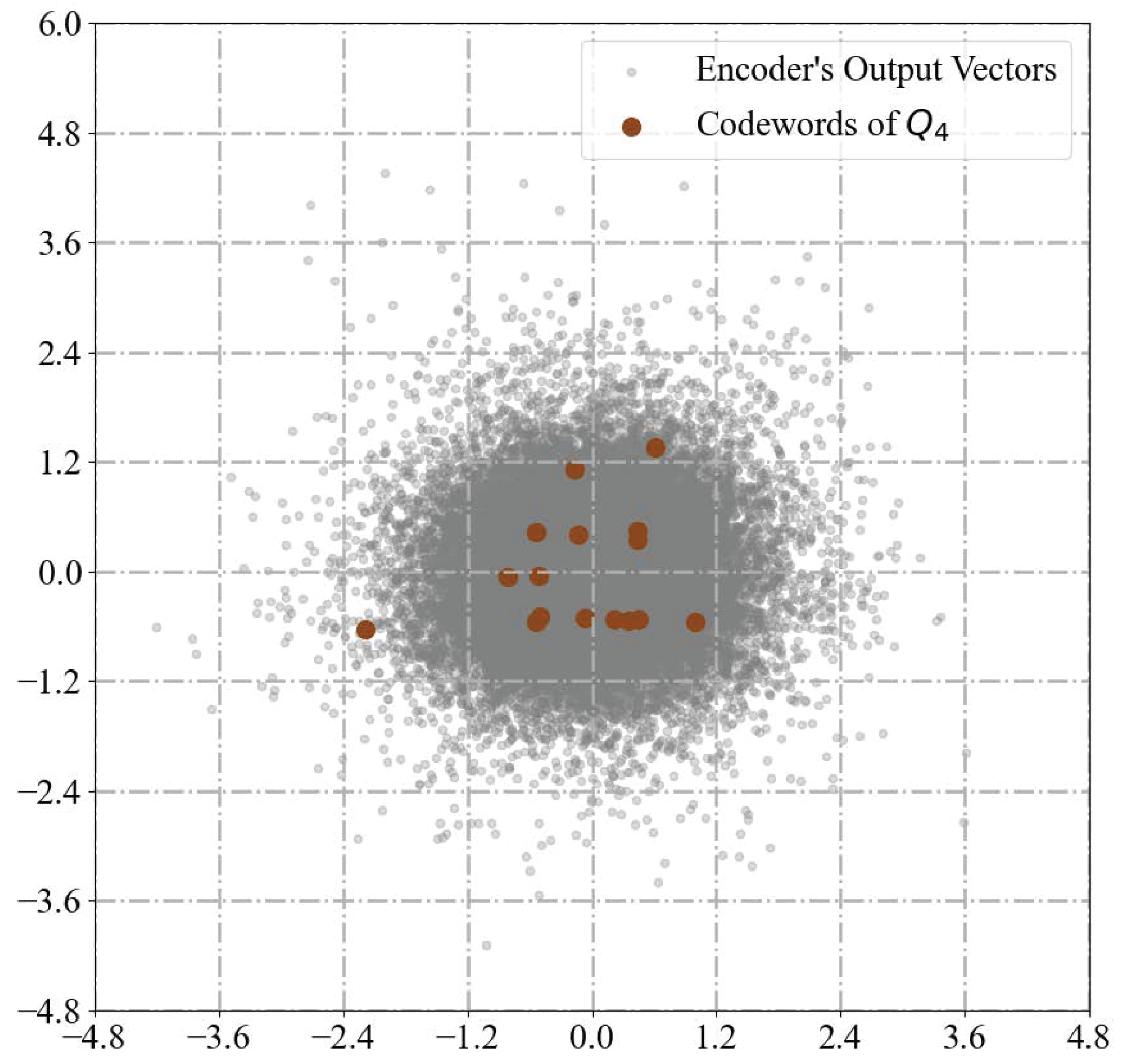}}
        \subfigure[Codebook $Q_{6}$.]{{\label{fig2c}}\includegraphics[width=0.245\linewidth]{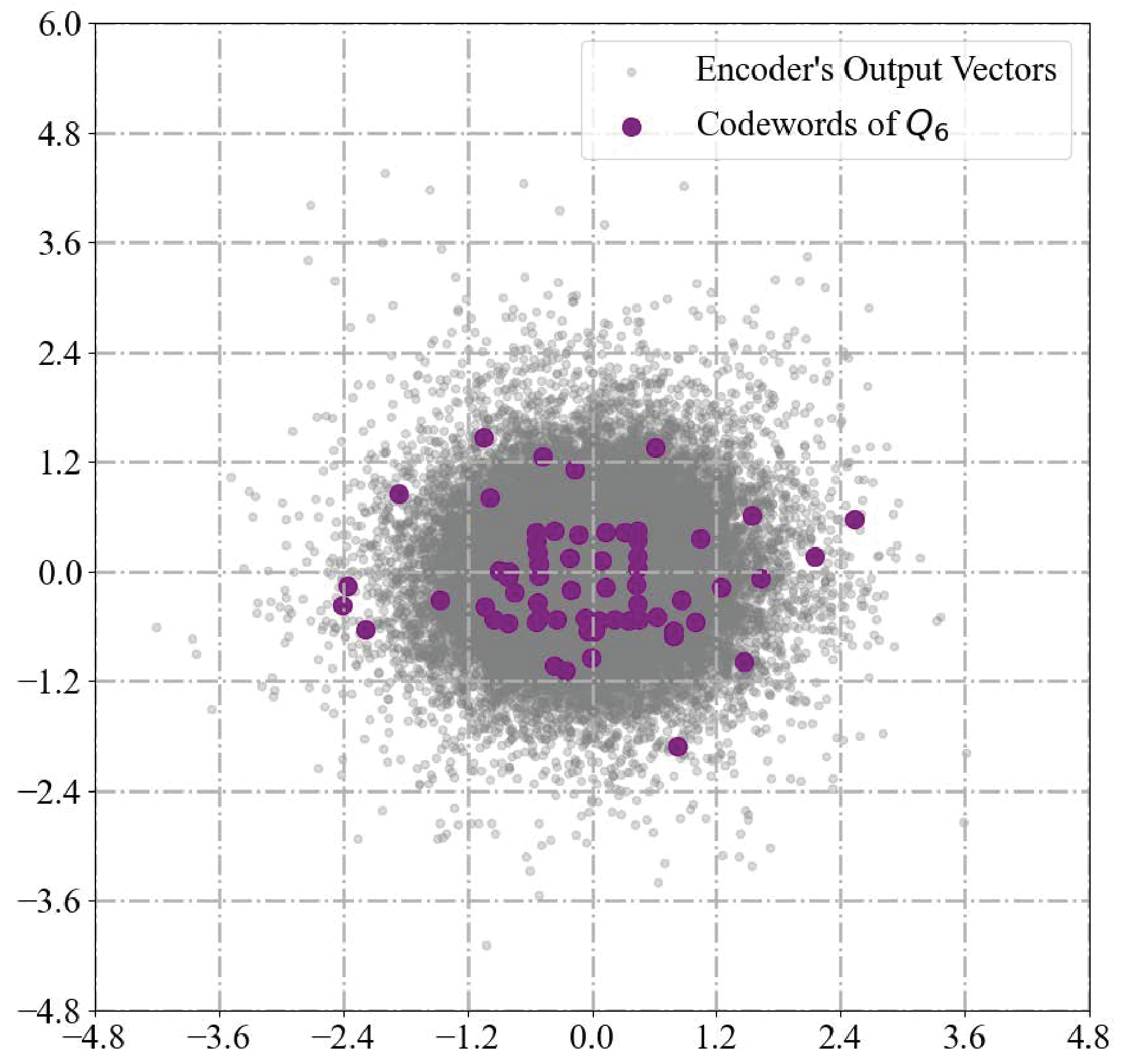}}
        \subfigure[Codebook $Q_{8}$.]{{\label{fig2c}}\includegraphics[width=0.245\linewidth]{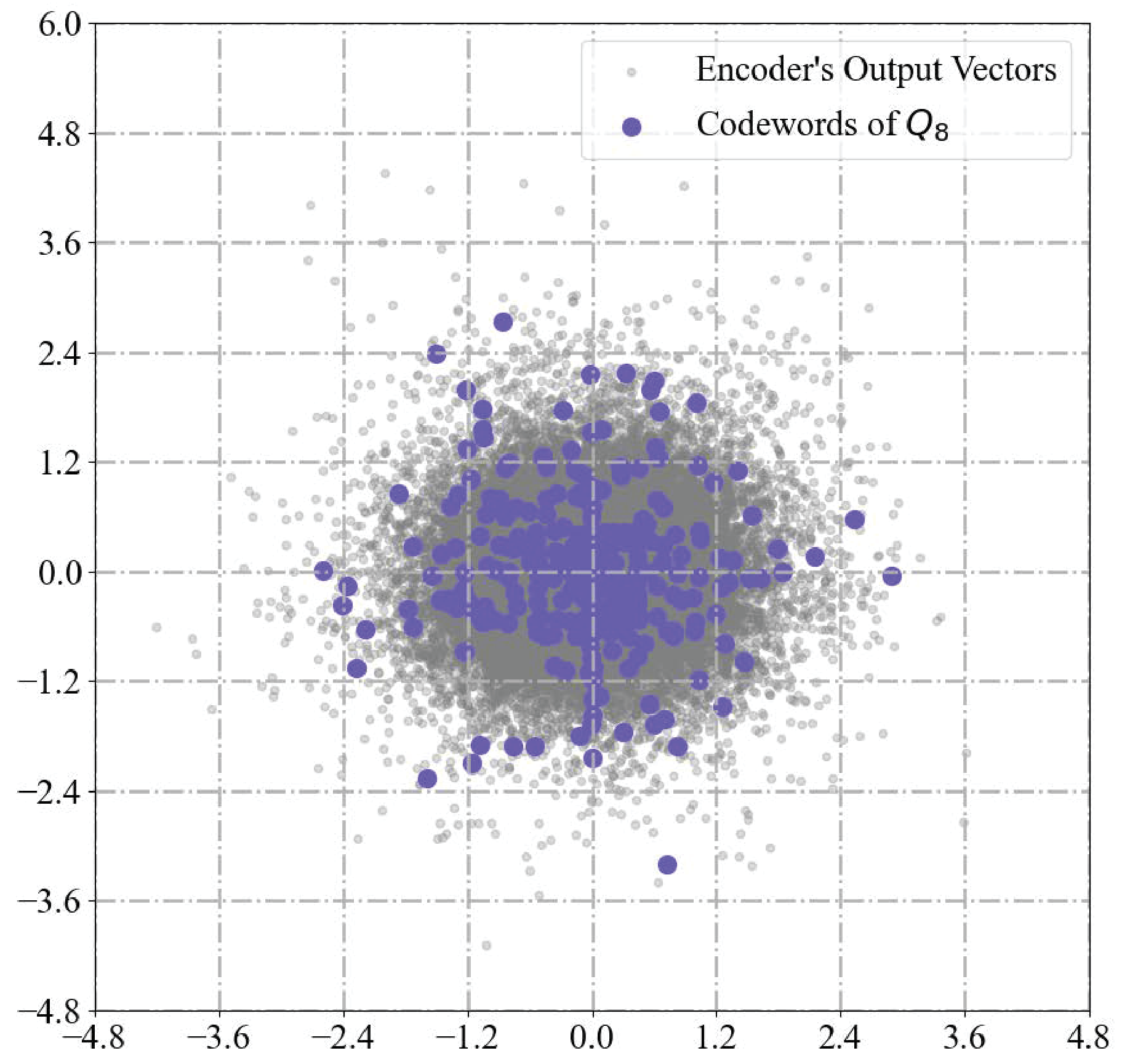}}
        \caption{Codebook size ${Q_{2}, Q_{4}, Q_{6}, Q_{8}}$ with $K_{\max}=256$.}
        \label{fig2}
\end{figure*}

\subsection{Progressive Learning Training Strategy}
\emph{Nested Codebook Training:} It is noteworthy that in our approach, adaptive capability can be achieved by combining nested encoding with progressive learning techniques. Unlike the single codebook, our training procedure follows the progressive learning approach, i.e., updating the codebook $Q_{l}$ along with the encoder and decoder based on the subsequent $Q_{l-1}$, with $Q_{0}=\emptyset$. For more details, when training at step $l \in\left\{1, \ldots, \log _{2} K_{\max}\right\}$, we already possess a codebook $Q_{l-1}$ along with the  trained encoder-decoder. Thus, when further expanding the codebook to $Q_{l}$, set the first $2^{l-1}$ codewords (i.e. $\{\boldsymbol{e}_{1}^{(l)}, \ldots, \boldsymbol{e}_{2^{l-1}}^{(l)}\}$) to have the same ordering as codewords in $Q_{l-1}$ (i.e. $\{\boldsymbol{e}_{1}^{(l-1)}, \ldots, \boldsymbol{e}_{2^{l-1}}^{(l-1)}\}$), while setting the remaining codewords to be random $D\times 1$ vectors.

In order to make the whole end-to-end system more adaptive to all sub-codewords of $Q_{l}$, we further use the corresponding loss function to train the MR-ToC framework. Taken together, this loss function not only ensures the inference performance about all codebooks in $Q_{l}$, but also encourages that the first $2^{l-1}$ codewords of $Q_{l}$ do not deviate too much from the already learnt code words in a progressive learning approach. Thus, at time $t$, the loss function at step $l$ is
\begin{equation}\label{eq12}
\begin{array}{l}
\mathcal{L}_{\text {MR-ToC}}^{(l)}\left(y_{t} ; \boldsymbol{x}_{t}\right)=\sum_{j=1}^{l} \mathcal{L}\left(y_{t} ; \hat{y}_{t}^{(j)}\right)+\left\|\mathbf{sg}\left[\boldsymbol{z}_{e,t}(\boldsymbol{x})\right]-\mathbf{e}_{j}\right\|_{2}^{2} \\
\\
\quad+\gamma_{j}\left\|\boldsymbol{z}_{e,t}(\boldsymbol{x})-\mathbf{sg}\left[\mathbf{e}_{j}\right]\right\|_{2}^{2}+\eta_{l} \sum_{k=1}^{2^{l-1}}\left\|\boldsymbol{e}_{k}^{(l)}-\boldsymbol{e}_{k}^{(l-1)}\right\|_{2}^{2},
\end{array}
\end{equation}
where the first term is the task-dependent loss, (e.g., cross entropy for classification), which considers the use of all sub-codebooks, thus encouraging the system to be rate-adaptive. The last term penalizes the learned codebook from deviating from the learned codewords so far. And this penalty coefficient $\eta_{l} \geq 0$, while setting $\eta_{1} \geq 0$ such that $\mathcal{L}_{\text {MR-ToC}}^{(1)}\left(y_{t} ; \boldsymbol{x}_{t}\right)$ coincides with the standard VQ loss term in (\ref{eq3}). Thus, the subsequent steps extend the codebook while preserving rate-adaptivity. The training algorithm is shown as Algorithm \ref{alg1}.

\emph{Complexity Analysis:}  The complexity of Algorithm \ref{alg1} is inherently dependent on the network architectures of the encoder and task inference modules. To illustrate, we consider the typical convolutional neural networks (CNNs). Let $F \times F$ represent the filter size, $C_{out}$ the number of filters, $C_{in}$ the number of input channels, and $W_1 \times W_2$ the feature map size. Assuming the neural network comprises $T$ convolutional layers, the complexity of Algorithm \ref{alg1}, in the context of progressively designing $\log_2 K_{\max}$ codebooks, can be expressed as $\mathcal{O}\left( TF^2 C_{in} C_{out} W_1 W_2\log_2 K_{\max}\right)$.

\section{Experiments and Discussions} 

In this section, we conduct experiments to demonstrate the superiority of the MR-ToC scheme, and choose a classification task as an example, without limitation.

\subsection{Experimental Settings}

To verify performance, we choose the CIFAR-10 dataset, which consists of 50000 train images and 10000 test images of size $32 \times 32$ pixels.
Then, we exploit several Inverted ResNets obtained by splitting the MobileNetV2 \cite{8578572} to form the learning backbones, which is designed to be suitable for hardware-limited edge device. The classification accuracy is used to measure the inference utility. 
In addition, we compare the proposed scheme with two baseline methods, i.e.,information-bottleneck-based single-rate ToC (IBSR-ToC) \cite{10159007} and cross-entropy-based single-rate ToC (CESR-ToC) \cite{2018NECST, 10418996}.
To ensure a fair comparison, we utilize the same neural network inputs and outputs across all local devices to address the same task, with consistent computational and memory constraints for all methods.
In the experiments, we set codebook size $K_{\max}=256$, the number of features $N_f=1000$, and the codeword dimensionality $D=2$ or $D=4$, corresponding numbers of sub-vectors $M=500$ and $M=250$. 
Without otherwise noted, we assume the channel model has the same error probability $\varepsilon_{train}$ in both the training phases.


\subsection{Experimental Results}

Then, to present progressive learning utility, we illustrate the learned nested codebooks $Q_{2}$, $Q_{4}$, $Q_{6}$, and $Q_{8}$, as well as the distribution of encoder’s outputs in Fig. \ref{fig2}. As can be seen in Fig. \ref{fig2}, the transceiver continuously interacts during end-to-end training, while our designed codebooks are constantly adapting to the dynamic channel conditions, and can gradually and accurately match the features extracted and mapped by the encoder. Moreover, when the rate level is gradually increased, the increased codewords are able to better capture the encoder vectors to perform the remote inference task.

\begin{figure}[t]
\centering
\includegraphics[scale=0.475]{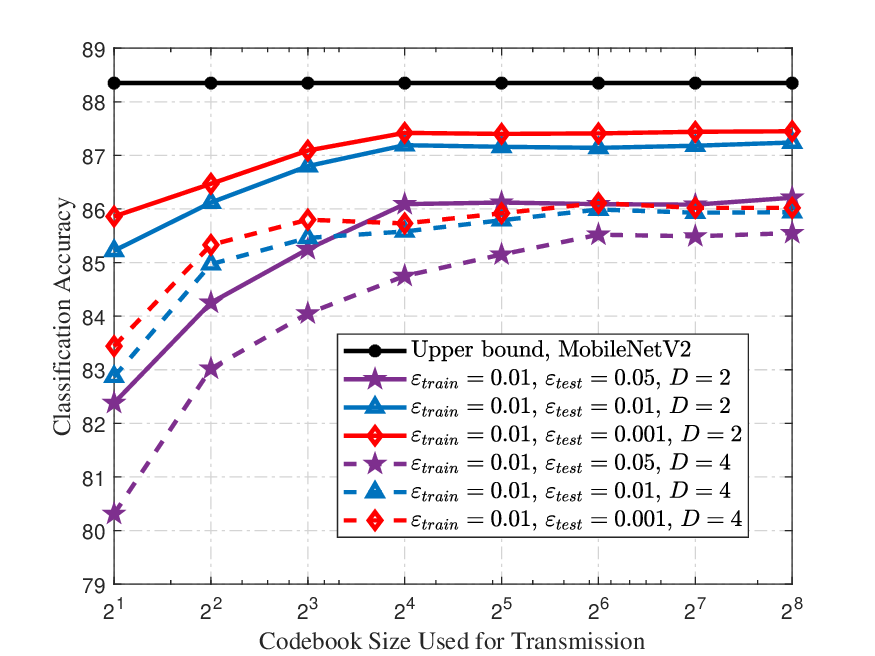}
\caption{Inference performance of MR-ToC over dynamic channels vs codebook size, where codewords dimension $D$ are set as $2$ and $4$.}
\label{fig3}
\end{figure}

In Fig. \ref{fig3}, we compare the MR-ToC performance under different channel conditions as well as MobileNetV2 without quantization. The MR-ToC scheme is trained under $\varepsilon_{ {train}}=0.01$ and tested under the selected samples $\varepsilon_{ {test}} \in\{0.001, 0.01, 0.05\}$, respectively. It can be observed that the MR-ToC performance gradually increases as the channel conditions become better in the same codeword dimension. As the codebook uses more bits, the accuracy correspondingly increases in each channel condition. Note that even in the worse channel conditions, the final performance approximates the case of good channel conditions as learned codewords increases. This is also due to the fact that MR-ToC co-adapts between the codewords of each sub-codeword to learn a better model based on the channel conditions. Acceptable inference performance can be achieved even with fewer bits transmitted in a regular channel. This highlights the effectiveness of our approach as it provides significant flexibility and substantial savings in memory and computational resources while adapting to channel variations.
Another comparison in  Fig. \ref{fig3}, we compare the codeword dimensions $D=4$ and $D=2$. The results show that comparing both codeword dimensions at the same training and test error rate level. That with $D=4$ is less accurate than that with $D=2$, by a decrease of about $1\%-1.5\%$. This is because the number of coding bits of $D=4$ is less than that of $D=2$ as $B=M\log_2 K$. This is also because it is more difficult to capture various features in $4$-D vectors than in $2$-D vectors in the same codebooks size.

\begin{figure}[t]
\centering
\includegraphics[scale=0.24]{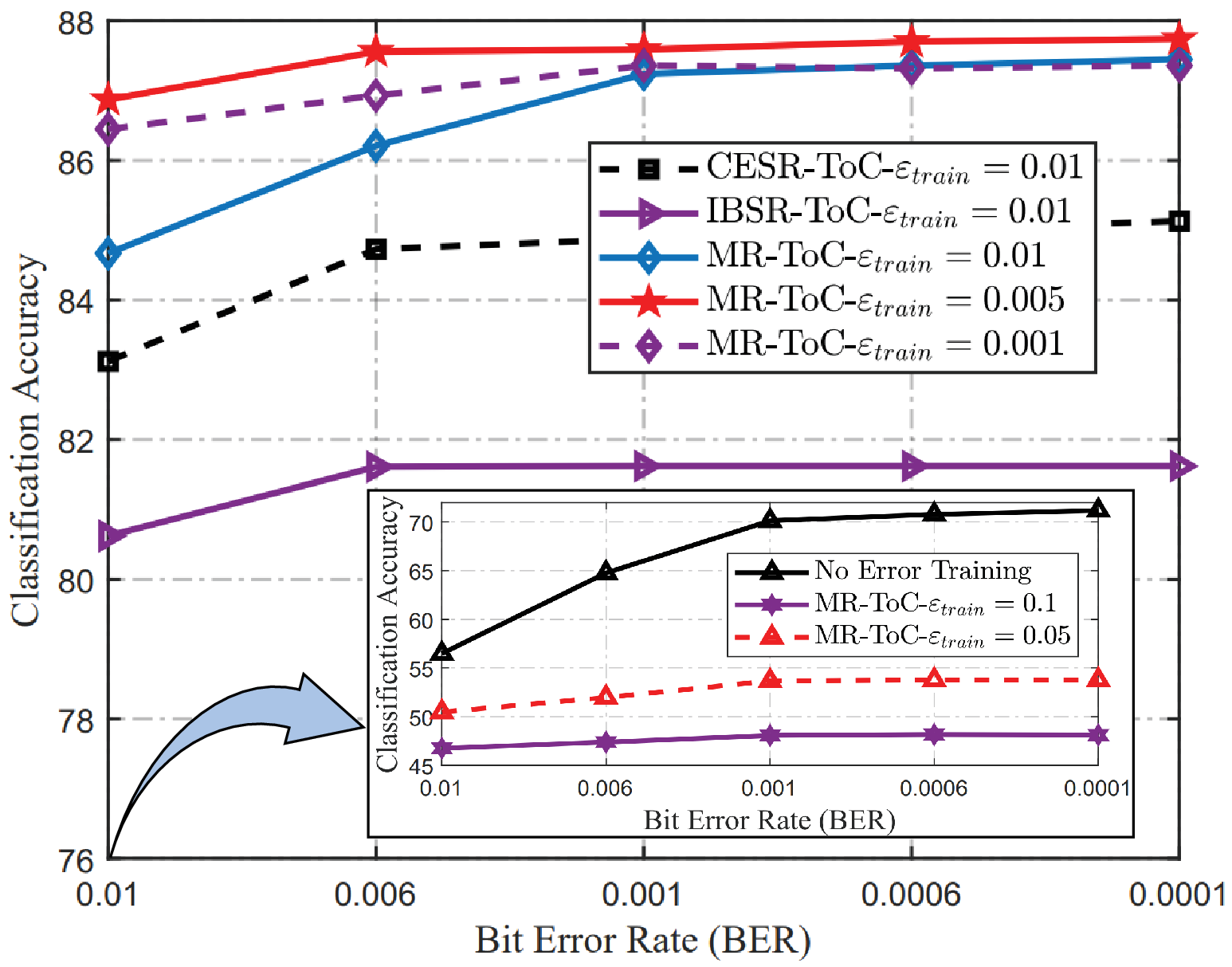}
\caption{Task inference performance of different ToC schemes over dynamic channels test under different BERs, where $D=2$.}
\label{fig4}
\end{figure}


In Fig. \ref{fig4}, we transform the different transmission error probabilities into the corresponding BER using equation (\ref{eq9}) as the horizontal axis. Then, we evaluate the performance of different ToC schemes under various transmission error rates ($\varepsilon_{train}$), and compare it with two state-of-the-art schemes. As observed, the performance deteriorates as the transmission error probability increases in each training condition. Notably, end-to-end joint training consistently yields superior inference performance (i.e., $86\%$–$88\%$) when conducted under low error rates (e.g., $\varepsilon_{train}={0.001, 0.005, 0.01}$) compared to training in error-free conditions. Conversely, satisfactory task performance (i.e., $50\%$) is not achieved when joint training is performed under conditions of high transmission error rates (e.g., $\varepsilon_{train}={0.05, 0.1}$). This indicates that selecting an appropriate transmission error probability is crucial for the effective training of MR-ToC. Additionally, when the baseline scenario is trained with a transmission error probability under standard channel conditions, MR-ToC demonstrates an accuracy improvement of approximately $2\%$–$5\%$ over the baseline scheme across various BERs. This result underscores the enhanced ability of MR-ToC to address multi-rate challenges and maintain inference task performance.


\section{Conclusion}
This letter proposed an MR-ToC for remote inference under dynamic channel conditions. The communication pipelines are conceptualized as multilevel symmetric discrete channels to simulate different channel conditions for the entire end-to-end network.  An efficient nested codebook was designed by the adaptive tuning scheme, which can learn the JSCC encoder and inference rules through a progressive learning approach. Experiment results demonstrate the proposed MR-ToC can provide good inference at multiple coding levels and can adapt to fluctuating communication environments. 

\bibliographystyle{IEEEtran} 
\bibliography{bib}

\end{document}